\documentclass[twocolumn,showpacs,preprintnumbers,amsmath,amssymb]{revtex4}
\usepackage{graphicx}
\usepackage{dcolumn}
\usepackage{bm}
\usepackage{amssymb}
\usepackage{amsmath}
\usepackage{dcolumn}

\RequirePackage{xspace}
\usepackage{relsize}
\def\babar{\mbox{\slshape B\kern-0.1em{\smaller A}\kern-0.1em
    B\kern-0.1em{\smaller A\kern-0.2em R}}}

\def\invfb   {\ensuremath{\mbox{\,fb}^{-1}}\xspace}
\newcommand{\gev}{\ensuremath{\mathrm{\,Ge\kern -0.1em V}}\xspace}
\newcommand{\mev}{\ensuremath{\mathrm{\,Me\kern -0.1em V}}\xspace}

\newcommand{\SLACPubNumber} {12784}

\begin{document}
{\pagestyle{empty}

\begin{flushleft}

SLAC-PUB-\SLACPubNumber  \\

\end{flushleft}

\title{
\Large \bf Extraction of \mbox{{\boldmath $|V_{ub}|$}} with Reduced Dependence on Shape Functions
}
\bigskip

\bigskip \bigskip

\begin{abstract}
\noindent
Using \babar\ measurements of the inclusive electron spectrum 
in $B \to X_u e \nu$ decays and the inclusive photon spectrum 
in $B \to X_s \gamma$ decays, we extract the magnitude of the 
CKM matrix element $V_{ub}$. 
The extraction is based on theoretical calculations designed to
reduce the theoretical uncertainties by exploiting the assumption that 
the leading shape functions are the same for all $b \to q$ transitions ($q$ is a 
light quark). The results agree well with the previous
analysis, have indeed smaller theoretical errors, but are presently limited by the
knowledge of the photon spectrum and the experimental errors on the
lepton spectrum.
\vfill

\end{abstract}

\author{V.~B.~Golubev}
\author{Yu.~I.~Skovpen}
\affiliation{Budker Institute of Nuclear Physics, Novosibirsk 630090, Russia }
\author{V.~G. L\"uth}
\affiliation{Stanford Linear Accelerator Center, Stanford, California 94309, USA }

\maketitle

}

\section{Introduction} 
\label{sec:introduction}

The determination of the magnitude
of the Cabibbo-Kobayashi-Maskawa matrix element
$V_{ub}$ from charmless semileptonic $B$  decays 
is complicated by the fact that over most of the phase space $B\to X_c\ell\nu$ 
decays dominate and are very difficult to distinguish from the 
signal $B\to X_u\ell\nu$ decays. Here $X_c$ and $X_u$ refer to hadrons,
mostly mesons, with and without charm.
To reduce impact of the dominant $B \to X_c \ell \nu$ background, partial rates for $B\to X_u\ell\nu$ 
decays are measured in regions of phase space where these background 
decays are forbidden or highly suppressed, for instance, 
near the endpoint of the energy spectrum of the charged lepton $\ell$. 
The partial decay rates are extrapolated to 
the total rate by comparing the experimentally measured rate with 
a theoretical prediction.  

Theoretically, the most precise predictions can be made for the
total $B\to X_u\ell\nu$ decay rate. Accounting for restriction in phase 
space is difficult because decay spectra close to the kinematic limit 
are susceptible to non-perturbative strong-interaction effects.
Theoretical tools for the calculations of the partial inclusive decay rates
are QCD factorization and local operator product expansions (OPE). These 
calculations separate non-perturbative from perturbative quantities, and 
use expansions in inverse powers of the $b$-quark mass, $m_b$, and 
in powers of the strong coupling $\alpha_s$~\cite{ope}.  
At leading order in $\Lambda_{\mathrm{QCD}}/m_b$, the non-perturbative bound-state 
effects are accounted for by a shape function describing the  
``Fermi-motion'' of the $b$ quark inside the $B$ meson.  These shape functions cannot
be calculated.  Different shapes have been proposed, and parameters
defining these shapes have to 
be extracted from data. This introduces significant additional hadronic 
uncertainties.  At leading order, these shape functions are assumed 
to be universal functions for $b \to q$ transitions, 
where $q$ is a light quark, either $s$ or $u$. 

In the past, we have extracted the non-perturbative parameters of 
these shape functions, the $b$-quark mass $m_b$ and its mean kinetic 
energy squared, $\mu^2_{\pi}(\mu)$, from the moments of the inclusive photon 
spectrum in $B\to X_s\gamma$ decays, as well as from moments of the hadron mass
and the lepton energy distributions in $B\to X_c\ell\nu$ decay.  
These parameters depend on 
the choice of the renormalization scale, $\mu$.

It was suggested many years ago~\cite{neubert1994} that $|V_{ub}|$ 
can be extracted with smaller theoretical uncertainties by combining integrals 
over the lepton spectrum in $B\to X_u\ell\nu$ decays with weighted 
integrals over the photon spectrum from $B \to X_s \gamma$ decays, each
above a cut-off energy $E_0$.  
The underlying assumption is that the QCD interactions affecting these two 
processes are the same and thus will cancel to first order in the appropriate 
ratio of weighted decay rates.  
The advantage of this approach is that it reduces the sensitivity 
to the choice of the shape-function parameterization and thus avoids 
uncertainties that are difficult to quantify.

In the following, we extract $|V_{ub}|$
using two different prescriptions, one proposed by Leibovich, Low, 
and Rothstein (LLR)~\cite{m3_llr1,m3_llr2, m3_llr_nr}, 
and the other based on more recent calculations by
Lange, Neubert, and Paz (LNP)~\cite{lnp_2loop,lange}.
The two prescriptions use different calculations of the 
weight functions, and thus result in different estimates of the 
theoretical uncertainties.

The \babar\ Collaboration was the first to apply the LLR prescription 
~\cite{m3_llr1,m3_llr2} to extract $|V_{ub}|$,
based on  measurements of the hadron mass spectrum in charmless
semileptonic $B$ decays and the inclusive photon spectrum~\cite{bbr_urs}.

\section{Experimental Input}

The experimental inputs for this analysis are the published
\babar\  measurements of the inclusive electron energy spectrum
in $B\to X_u e\nu$ decays~\cite{bbr_xulnu}, and of the inclusive
photon-energy spectrum in $B\to X_s\gamma$ 
decays~\cite{bbr_bsgamma}.  These measurements are based on a data
sample corresponding to a total integrated luminosity of about 80~\invfb.
The measured spectra will be integrated above an energy $E_0$, measured
in the $B$-meson rest frame. 

\subsection{Inclusive Lepton Spectrum in {\boldmath $B\to X_ue\nu$} Decays}
\begin{figure}
\begin{center}
{\includegraphics[height=7.5cm]{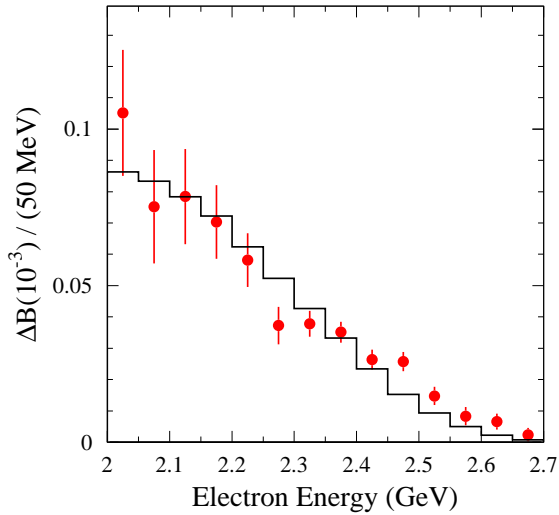}
}
\caption{
The differential branching fraction for 
$B\to X_u e \nu$ decays~\cite{bbr_xulnu} as a function of the charged lepton 
momentum, measured in the $\Upsilon(4S)$ rest frame.  The error bars indicate
the statistical errors; the histogram represents the theoretical
prediction~\cite{neubert_n1}. 
}
\label{f:electron}
\end{center}
\end{figure}

The inclusive electron-energy spectrum above 2.0\gev,  
measured in the 
$\Upsilon(4S)$ rest frame, is shown in Fig.\ref{f:electron}.
The data are fully corrected for detection efficiencies as well as
final state radiation and bremsstrahlung.  They are
normalized to the total number of charged and neutral $B$-meson decays and
are presented as differential branching fraction.

For the extraction of $|V_{ub}|$ we need to transform the measured
partial branching fraction from the $\Upsilon(4S)$ rest frame to the 
branching fraction in the $B$-meson rest frame, integrated over 
the spectrum above an energy cut-off at $E_0$ .  
This is done using correction factors
derived from the predicted electron spectrum \cite{neubert_n1},
calculated with the shape-function parameters determined by a global
fit~\cite{babarcf}
to moments of inclusive distributions in semileptonic and radiative
$B$-meson decays.
The systematic uncertainties for this transformation are estimated 
by varying the shape-function parameters within their experimental
uncertainties.
The resulting partial branching fractions for different values of the 
electron-energy cut-off, $E_0$, measured in the $\Upsilon(4S)$ 
and the $B$-meson rest frames, and correction factors relating 
the two, are listed in Table~\ref{t:brsl}.

\begin{table}[h]
\caption{Partial branching fractions for $B\to X_u e\nu$ decays
in units $10^{-3}$, integrated over the energy range from $E_0$ to
$2.6\,\mathrm{GeV}/c$, both in the $\Upsilon(4S)$ rest frame and the 
$B$-meson rest frame, as well as the correction factor relating the two.
The uncertainty of the correction factor 
reflects the uncertainty in the assumed shape of the spectrum.
}

\begin{center}
\begin{ruledtabular}
\begin{tabular}{cccc}
$E_0$  & ${\Delta\cal B}(E_0)\cdot 10^3$ & Correction & $\Delta{\cal B}(E_0)\cdot 10^3$ \\
$[\mathrm{GeV}]$ & $ \Upsilon(4S)$ rest frame & factor & $B$ rest frame \\ \hline
$2.0$ & $0.572 \pm 0.041 \pm 0.065$ &$1.002 \pm 0.005$ &  $0.573\pm 0.077$ \\ 
$2.1$ & $0.392 \pm 0.023 \pm 0.038$ &$0.994 \pm 0.008$ &  $0.390\pm 0.044$ \\ 
$2.2$ & $0.243 \pm 0.011 \pm 0.020$ &$0.973 \pm 0.016$ &  $0.236\pm 0.023$ \\ 
$2.3$ & $0.148 \pm 0.006 \pm 0.010$ &$0.915^{+0.041}_{-0.034}$ & $0.135 \pm 0.012$ \\ 
$2.4$ & $0.075 \pm 0.004 \pm 0.006$ &$0.772^{+0.075}_{-0.084}$ & $0.058 \pm 0.008$ \\ 
\end{tabular}
\end{ruledtabular}
\end{center}
\label{t:brsl}
\end{table}

\subsection{Inclusive Photon Spectrum in {\boldmath $B\to X_s\gamma$} Decays}

\begin{figure}[h]
\begin{center}
{\includegraphics[height=6.4cm]{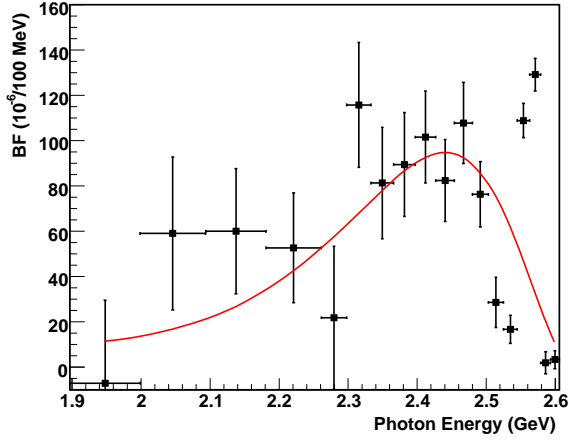}
}
\caption{
The differential branching fraction of inclusive $B\to X_s \gamma$ decays 
as a function of the photon energy in the $B$ rest frame.
The error bars indicate the statistical errors.  The data are compared to 
the theoretical prediction~\cite{neubert_n1} using shape function
parameters $m_b=4.67\;\mathrm{GeV}$, $\mu_\pi^2=0.16\;\mathrm{GeV}^2$
that have been 
obtained from the fit to $E_\gamma$ spectrum~\cite{bbr_bsgamma}. 
}
\label{f:bsg}
\end{center}
\end{figure}

The inclusive photon-energy spectrum~\cite{bbr_bsgamma} in $B\to X_s\gamma$ decays,
above $1.9\,\mathrm{GeV}$, measured in the B-meson rest frame, is shown in Fig.~\ref{f:bsg}.
This spectrum is measured as the sum of the photon spectra in 38 exclusive decay modes. 
The data are fully corrected for detection efficiencies and energy resolution.
They are normalized to the total number of charged and neutral
$B$-meson decays and presented as differential branching fractions.

\section{Theoretical Framework for the Extraction of {\boldmath $|V_{ub}|$} }

\subsection{Method I by Leibovich, Low, and Rothstein}
A. K.~Leibovich, I.~Low, and I. Z.~Rothstein~\cite{m3_llr1,m3_llr2} proposed 
a method for the extraction of the ratio $|V_{ub}|^2/|V_{tb}V^{*}_{ts}|^2$ 
without invoking knowledge of the shape function. 
Their calculation relates
$|V_{ub}|^2/|V_{tb}V_{ts}^*|^2$ to the experimentally
measured differential branching fractions for $B\to X_u \ell\nu$ and
$B\to X_{s}\gamma$ decays,

\begin{eqnarray}
\frac{\textstyle |V_{ub}|^2}{\textstyle |V_{tb}V^{*}_{ts}|^2} = 
\frac{\textstyle 3 \alpha |C^{(0)}_7(m_b)|^2}
{\textstyle \pi} (1+H^{\gamma}_{\mathrm{mix}})\times \nonumber \\ 
\bigg(\int_{x_B^c}^{1} dx_B \frac{\textstyle  d\Gamma}{\textstyle dx_B}\bigg ) \bigg / 
\bigg(\int_{x_B^c/\rho}^{1} du_B 
\frac{\textstyle d\Gamma^{\gamma}}{\textstyle du_B} 
w(u_B,x_B^c,\rho)\bigg).
\end{eqnarray}
\noindent 
The integration variables are
\begin{equation*}
x_B=\textstyle 2 E_l / \textstyle M_B,\; \; \; 
u_B=\textstyle 2 E_\gamma / \textstyle M_B,
\end{equation*} 
\noindent
with limits $x_B^c$ and $x_B^c/\rho$; $M_B$ refers to the $B$-meson mass.
The weight function is approximately linear as a function of $u_B$,
\begin{eqnarray}
w(u_B,x_B^c,\rho) = \nonumber \\
u_B^2 \int_{x_B^c}^{\rho u_B} dx_B 
K\Bigl(x_B; \frac{\textstyle 4}{\textstyle 3 \pi \beta_0} 
\ln (1 - \alpha_s \beta_0 l_{x_B/u_B})\Bigr), 
\end{eqnarray} 
\noindent and
\begin{eqnarray}
H^{\gamma}_{\mathrm{mix}} &=& \frac{\textstyle \alpha_s}{\textstyle 2 \pi C^{(0)}_7} 
\bigg[C^{(1)}_7 + C^{(0)}_2 \Re (r_2) + C^{(0)}_8 
\bigg(\frac{\textstyle 44}{\textstyle 9} - 
\frac{\textstyle 8\pi^2}{\textstyle 27}\bigg)\bigg], \nonumber \\
\Re(r_2) &\approx& -4.092+12.78(m_c/m_b-0.29),  \nonumber \\
K(x;y) &=& 6 \bigg\{ \bigg[1+\frac{\textstyle 4 \alpha_s}{\textstyle 3 \pi} 
(1-\psi^{\prime}(4+y))\bigg] 
\frac{\textstyle 1}{\textstyle (y+2)(y+3)} \nonumber \\
&-& \frac{\textstyle \alpha_s}{\textstyle 3 \pi} 
\bigg[\frac{\textstyle 1}{\textstyle (y+2)^2} - 
\frac{\textstyle 7}{\textstyle (y+3)^2} \bigg] \nonumber \\
&-& \frac{\textstyle 4 \alpha_s}{\textstyle 3 \pi} 
\bigg[\frac{\textstyle 1}{\textstyle (y+2)^3} -
\frac{\textstyle 1}{\textstyle (y+3)^3} \bigg]
\bigg\} - 3(1-x)^2, \nonumber \\
\psi(z) &=& \frac{\textstyle 1}{\textstyle \Gamma(z)}
\frac{\textstyle d}{\textstyle dz} \Gamma(z), \; \;
l_{x/u}=-\ln[-\ln(x/u)]. \nonumber 
\end{eqnarray} 
\noindent
The Wilson coefficients $C^{(0)}_7(m_b)$ , $C^{(0)}_8(m_b)$, {\it etc}.,
computed in NLO, can be found in \cite{kn_0,kn_3}.
The argument of $K$, $y=\frac{\textstyle 4}{\textstyle 3 \pi \beta_0} 
\ln (1 - \alpha_s \beta_0 l_{x_B/u_B})$, diverges for  $\alpha_s \beta_0 l_{x_B/u_B}=1$.
To avoid this pole, the integration limits are set to $x_B \leq \rho \; u_B$
 with $\rho < 0.999$.  For smaller values of $\rho$, the weight function 
deviates and thus the extracted value of $|V_{ub}|$ changes, and the 
uncertainty of this scheme increases.

To check the impact of the resummation of the Sudakov logarithms we
also tried a non-resummed version of the LLR weight function~\cite{m3_llr_nr}:
\begin{eqnarray}
w(u_B,x_B^c) &=& u_B^2 \int_{x_B^c}^{u_B}dx_B\bigg(1-3(1-x_B)^2 \nonumber \\
&+& \frac{\alpha_s}{\pi}
(\frac{7}{2}-\frac{2\pi^2}{9}-\frac{10}{9}\ln(1-\frac{x_B}{u_B}))\bigg).
\end{eqnarray} 

\begin{figure}
\begin{center}
{\includegraphics[height=5.2cm]{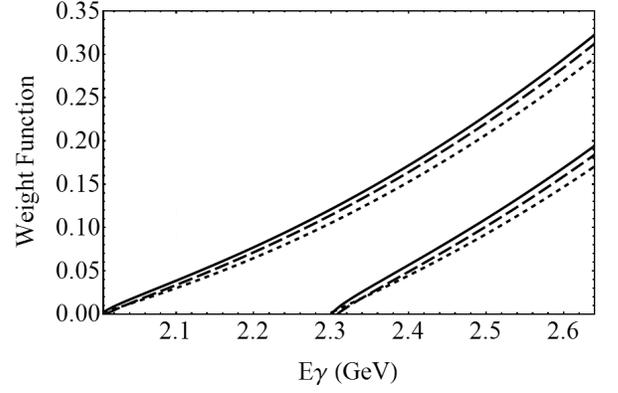}}
\caption{
The weight functions for Method I (LLR) for two values of
$E_0=2.0\,\mathrm{GeV}$ and $E_0=2.3\,\mathrm{GeV}$,
corresponding to different values of $\rho=0.9983$
(solid lines), $\rho=0.995$ (dashed lines), and non-resummed weight function
(dotted lines).
}
\label{f:llr_wf}
\end{center}
\end{figure}

\noindent
This calculation (Eq.~$1 - 3$) includes NLO perturbative corrections, but
does not take into account power corrections. Estimates of the 
theoretical uncertainties are discussed in Section~4.
The weight function is shown in
Fig.~\ref{f:llr_wf},  for $E_0=2.0\,\mathrm{GeV}$ and $E_0=2.3\,\mathrm{GeV}$
and different values of the parameter $\rho$.

M. Neubert~\cite{neubert2001} performed calculations that are quite similar
and give results that are very close to the 
ones obtained for this method.

\subsection{Method II by Lange, Neubert, and Paz}

The second method for the extraction of $|V_{ub}|$ is an application of 
the more recent two-loop calculations by 
B.~Lange, M.~Neubert, and G.~Paz~\cite{lnp_2loop}, and by B. Lange~\cite{lange},
performed for the high end of the lepton and photon energy spectra in 
$B\to X_u \ell \nu$ and $B\to X_s \gamma$ decays.

Unlike the previously described methods, this one relates $|V_{ub}|$
to the measured partial $B\to X_u \ell\nu$ branching fraction and normalized
photon spectrum in the $B\to X_s\gamma$ decay,
\begin{eqnarray}
|V_{ub}|^2&=& \frac{\frac{1}{\tau_B} \int\limits_{E_0}^{M_B/2} dE_{\ell}\, (dB(B\to X_u \ell\nu)/dE_{\ell})}
{\int\limits_{E_0}^{M_B/2} dE_\gamma\, w(E_\gamma, E_0)S(E_\gamma)+\Gamma_{\mathrm{rhc}}(E_0)}, \\
S(E_\gamma)&=&\frac{\textstyle 1}
{\textstyle \Gamma(B \to X_s \gamma)_{E_{\gamma}>E_{\mathrm{min}}}}
\frac{\textstyle d\Gamma(B \to X_s \gamma)}
{\textstyle dE_\gamma},
\end{eqnarray} 

\noindent where $ \Gamma_{\mathrm{rhc}}(E_0)$ represents residual hadronic 
power corrections, which in \cite{lnp_2loop} were absorbed into the 
weight function. $E_{\mathrm{min}}=1.90\gev$ is chosen as the lower limit for
the normalization of the photon energy spectrum; the corresponding branching fraction is $3.95 \times 10^{-4}$.

\begin{figure}
\begin{center}
{\includegraphics[height=5.2cm]{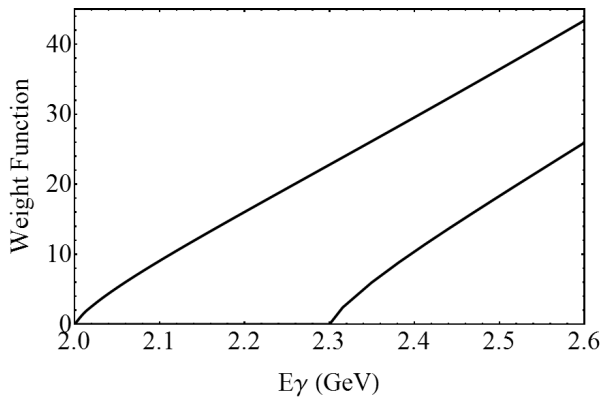}}
\caption{
The weight function (in units of $\mathrm{ps}^{-1}$) for Method II (LNP),
for $E_0=2.0\,\mathrm{GeV}$ (upper curve) and $E_0=2.3\,\mathrm{GeV}$ (lower curve).
}
\label{f:bl_wf}
\end{center}
\end{figure}

This calculation contains perturbative corrections at
NNLO at the so-called ``jet scale'', $\mu_i\sim\sqrt{m_b\Lambda_{\mathrm{QCD}}}$,
and at NLO at the so-called ``hard scale'', $\mu_h\sim m_b$. Also included
are the first-order power
corrections, which are separated into two parts: kinematic corrections
and residual hadronic corrections. The kinematic corrections do not
introduce hadronic uncertainties and are applied directly to the
weight function. The residual hadronic corrections include 
subleading shape functions for which the functional form is unknown. This
introduces significant theoretical uncertainties. 
The weight functions, calculated for $E_0=2.0\,\mathrm{GeV}$ and
$E_0=2.3\,\mathrm{GeV}$, are shown in Fig~\ref{f:bl_wf}.

\section{Determination of {\boldmath $|V_{ub}|$}}
\subsection{Calculation of weighted integrals of $B\to X_s\gamma$ photon spectrum}

For each of the two methods we extract $|V_{ub}|$ from a ratio of the
$B\to X_u l\nu$ 
partial branching fraction and a weighted integral over the photon
spectrum for $B \to X_s \gamma$.
Each method introduces a specific weight
function $w(E_\gamma,E_0)$. The value of the weighted integral of the 
photon spectrum above a minimum energy $E_0$,
\begin{equation}
I(E_0)=\int\limits_{E_0}^{M_B/2} dE_\gamma\, w(E_\gamma, E_0)
 d\Gamma(E_\gamma)/dE_\gamma ,
\end{equation}
is taken as a sum over bins in the photon energy in the $B$-rest frame,
\begin{equation}
\tilde{I}(E_0)=\sum_i w(E_{\gamma \,i}, E_0)
 \Bigl (\frac{d\Gamma(E_\gamma)}{{dE_\gamma}}\Bigr )_i \Delta E_{\gamma\, i}.
\end{equation}

The uncertainty for this sum is estimated using the standard error propagation 
for a linear combination of random variables:
\begin{equation}
\sigma^2=\sum_{ij} w(E_{\gamma\,i}, E_0) w(E_{\gamma\,j}, E_0)  
\Delta E_{\gamma\,i} \Delta E_{\gamma\,j} V_{ij},
\end{equation}
\noindent where $V_{ij}$ is a covariance matrix of the differential rate
$d\Gamma(E_\gamma)/dE_\gamma$  for different photon energy bins.
Assuming uncorrelated statistical errors and correlated systematic errors with
known correlation matrix $R_{ij}$, the covariance matrix is of the form
\begin{equation}
V_{ij} = \sigma_{\mathrm{stat}}^i \sigma_{\mathrm{stat}}^j \delta_{ij} +
\sigma_{\mathrm{syst}}^i \sigma_{\mathrm{syst}}^j R_{ij}.
\end{equation}
The correlation matrix $R_{ij}$ was provided by the \babar\ Collaboration~\cite{bbr_bsgamma}.

\subsection{Results on {\boldmath $|V_{ub}|$} and Error Estimation}

\subsubsection{Method I (LLR)}
The results of the extraction of ${|V_{ub}|}/{|V_{tb}V^{*}_{ts}|}$ based on 
Method I are presented in Table~\ref{t:m1}.
The theoretical uncertainties of the NLO calculations 
have been estimated~\cite{m3_llr1,m3_llr2} 
to be ${\cal O}(\Lambda_{\mathrm{QCD}}/M_B)$, resulting in a relative error of 
about 6\% for ${|V_{ub}|}/{|V_{tb}V^{*}_{ts}|}$. The extraction of
${|V_{ub}|}/{|V_{tb}V^{*}_{ts}|}$ was done with resummed Sudakov logarithms.
As a cross check, the extraction was also performed without resummation,
resulting in decreases of the weight function by $\sim 6\%$ at
$E_0=2.0\,\mathrm{GeV}$ and by up to $\sim 12\%$ at $2.4\,\mathrm{GeV}$.

\begin{table*}[floatfix]
\caption{The results for 
${|V_{ub}|}/{|V_{tb}V^{*}_{ts}|}$ and $|V_{ub}|$
for Method I (LLR). The extraction uses the weight functions
based on resummed Sudakov logarithms with $\rho=0.9983$.
The first error represents the error from the measured $B \to X_u e \nu$ 
partial branching fraction, the
second error is from the measured $B\to X_s \gamma$ spectrum, and the 
third is the estimated theoretical uncertainty.  The fourth error on $|V_{ub}|$ is from
the $|V_{ts}|$ uncertainty.
}
\begin{center}
\begin{ruledtabular}
\begin{tabular*}{15cm}{ccc}
$E_0$ [GeV]& 
${|V_{ub}|}/{|V_{tb}V^{*}_{ts}|}$ & $ |V_{ub}| \cdot 10^3 $ \\ \hline
$2.0$ & $0.106 \pm 0.007 \pm 0.007 \pm 0.006$ & $4.28 \pm 0.29 \pm 0.29 \pm 0.26 \pm 0.28 $ \\
$2.1$ & $0.100 \pm 0.006 \pm 0.006 \pm 0.006$ & $4.06 \pm 0.23 \pm 0.25 \pm 0.24 \pm 0.27 $ \\
$2.2$ & $0.093 \pm 0.005 \pm 0.005 \pm 0.006$ & $3.78 \pm 0.18 \pm 0.21 \pm 0.23 \pm 0.25 $ \\
$2.3$ & $0.091 \pm 0.004 \pm 0.005 \pm 0.006$ & $3.69 \pm 0.16 \pm 0.19 \pm 0.22 \pm 0.25 $ \\
$2.4$ & $0.090 \pm 0.006 \pm 0.004 \pm 0.006$ & $3.64 \pm 0.25 \pm 0.17 \pm 0.22 \pm 0.24 $ \\
\end{tabular*}
\end{ruledtabular}
\end{center}
\label{t:m1}
\end{table*}

To translate these results into $|V_{ub}|$ we exploit the constraints of
the unitarity
of the CKM matrix resulting in
$|V_{tb}|\cong 1+{\cal O}(\lambda^4)$, where $\lambda=0.226$ is the sine
of the Cabibbo angle. Using 
$|V_{ts}|=(40.6 \pm 2.7)\cdot 10^{-3}$~\cite{PDG_2006},  we 
calculate $|V_{ub}|$ (see Table~\ref{t:m1}).

The results show high stability with respect to variations of the
lepton energy cut in the range from 2.0 to 2.4~GeV. The  partial charmless
semileptonic branching fraction decreases in this range from 25\% to 2.3\%
of the total $B\to X_u l\nu$ branching fraction \cite{bbr_xulnu}. The observed
stability is quite surprising because near the endpoint of the lepton energy
spectrum the theoretical calculations are expected to break down,  and this
should lead to increasing theoretical uncertainties.

\subsubsection{Method II (LNP)}

The results on the extraction of $|V_{ub}|$ based on Method II
are presented in Table~\ref{t:m4}. The estimated individual theoretical
uncertainties~\cite{lange} are added in quadrature to obtain the total theoretical error.
The first order non-perturbative hadronic power corrections,
not considered in Method I, are split into two contributions: 
kinematic corrections, which are included in the weight function and
depend on the scale for the calculation of kinematic power 
corrections, $\overline{\mu}$, but are independent of intermediate scale,
$\mu_i$, the hard scale, $\mu_h$, and residual hadronic corrections.  
These corrections
depend on unknown subleading shape functions,
which affect the $B\to X_u l\nu$ and $B\to X_s\gamma$ spectra in different ways 
and thus do not cancel in the weighted ratio in the integrals.
This uncertainty, $\sigma_{\mathrm{had}}$, is estimated as suggested by
Lange~\cite{lange}, namely by varying the shape and the parameters 
of the subleading shape functions.

The error $\sigma_{\mathrm{pert}}$ is the uncertainty of the NNLO approximation of the
shape function and the LO approximation of the power corrections. 
It was estimated from the dependence of the 
sum of the weighted integral over the photon
spectrum and residual hadronic correction on the hard scale,
$\mu_h=m_b/\sqrt{2}$, on the 
intermediate scale, $\mu_i=1.5\,\mathrm{GeV}$, and on
the scale for the power corrections, 
$\overline{\mu}=1.5\,\mathrm{GeV}$.
All three scales are varied by factors of $\sqrt{2}$ and $1/\sqrt{2}$
relative to their default values, and the largest variation of the
weighted integral is taken as the estimate of the perturbative uncertainty.

The errors $\sigma_{\mathrm{m_b}}$ and $\sigma_{\mathrm{parm}}$ are due
to uncertainties of the parameters that are
inputs to the calculation and were varied within their stated errors:
$m_b=4.61\pm 0.06$ \gev and $m_c/m_b=0.222\pm 0.027$, 
$m_s=90\pm 25$\mev. 
Here $m_b$ and $\mu_\pi^2$ are defined in the shape-function scheme
at a scale $\mu_*=1.5\,\mathrm{GeV}$, $m_s$ and the ratio $m_c/m_b$ 
are evaluated in the $\overline{MS}$
scheme at 1.5~GeV, where the ratio is scale-invariant.
The uncertainties on $\lambda_2=0.12\,\mathrm{GeV}^2$ and
$\mu_\pi^2=0.25\pm 0.10\,\mathrm{GeV}^2$ only enter into the subleading
shape functions, for which the uncertainties are assessed separately.

The error  $\sigma_{\mathrm{norm}}$ represents the uncertainty of the
normalization of the photon spectrum, which is estimated to be
about~6\%~\cite{lange}.

\begin{table*}[floatfix]
\caption{
The results for $|V_{ub}|$ based on Method II (LNP). 
The first error represents the error from the measured $B \to X_u e \nu$ 
partial branching fraction, and the second from the measured 
$B\to X_s \gamma$ spectrum. The contributions to the 
theoretical error, as described in the text,  are given together
with the total theoretical error.
}
\begin{center}
\begin{ruledtabular}
\begin{tabular}{cccccccc}
$E_0$ [GeV] & $|V_{ub}|\cdot 10^3 $ 
& $\sigma_{\mathrm{had}}\cdot 10^3 $ & $\sigma_{\mathrm{pert}}\cdot 10^3$ 
& $\sigma_{m_b}\cdot 10^3 $  & $\sigma_{\mathrm{parm}} \cdot 10^3$ 
& $\sigma_{\mathrm{norm}} \cdot 10^3$ & $\sigma_{\mathrm{theory}} \cdot 10^3$ \\ \hline
\rule[-0mm]{0mm}{4mm}
$2.0$ & $4.40 \pm 0.30 \pm 0.41$ & $\;^{+0.08}_{-0.07} $ & $ \;^{+0.03}_{-0.01}$ 
& $ \;^{+0.13}_{-0.12}$ & $\;^{+0.01}_{-0.00}$ & $ \pm 0.17 $ & $\;^{+0.23}_{-0.22}$ \\
\rule[-0mm]{0mm}{4mm}
$2.1$ & $4.55 \pm 0.26 \pm 0.45$ & $\;^{+0.15}_{-0.13} $ & $ \;^{+0.11}_{-0.05}$ 
& $ \;^{+0.16}_{-0.15}$ & $\pm 0.01 $ & $\pm 0.21$ & $\;^{+0.32}_{-0.29}$ \\
\rule[-0mm]{0mm}{4mm}
$2.2$ & $5.01 \pm 0.24 \pm 0.60$ & $\;^{+0.42}_{-0.33} $ & $ \;^{+0.40}_{-0.21}$ 
& $ \;^{+0.25}_{-0.22}$ & $\pm 0.01 $ & $\pm 0.32$ & $\;^{+0.71}_{-0.55}$ \\
\rule[-0mm]{0mm}{4mm}
$2.3$ & $6.99 \pm 0.31 \pm 1.60$ & $\;^{+3.90}_{-1.44} $ & $ \;^{+3.02}_{-1.00}$ 
& $ \;^{+0.75}_{-0.58}$ & $\;^{+0.04}_{-0.03}$ & $\pm 0.92$ & $\;^{+5.07}_{-2.05}$ \\
\end{tabular}
\end{ruledtabular}
\end{center}
\label{t:m4}
\end{table*}

We observe in Table~\ref{t:m4} a significant increase in the extracted
value of $|V_{ub}|$ for  $E_0 \ge 2.2$~\gev, and the theoretical 
uncertainties also show a rapid growth for $E_0>2.1\,\mathrm{GeV}$,
reaching 100\% above $E_0\simeq 2.3\gev$.
This is due to the power correction $\Gamma_{\mathrm{rhc}}(E_0)$ in the denominator
of the Eq.4. This correction is negative 
and almost independent of $E_0$. As $E_0$ increases,  the integral over the 
photon spectrum decreases and contribution from the $\Gamma_{\mathrm{rhc}}(E_0)$
to the total uncertainty increases.
This is not unexpected, because the effect of 
non-perturbative hadronic corrections increases in the region close to 
kinematic endpoints for both decays.

\section{Conclusion}

We have extracted the CKM matrix element $|V_{ub}|$ using 
published \babar\ measurements of the inclusive
lepton spectrum in $B\to X_u e\nu$ decays and inclusive photon spectrum in
$B\to X_s\gamma$ decays. By using the ratios of the weighted spectra 
for these two decays, the results are 
expected to be less model dependent than previous measurements 
relying on the extraction of the shape functions from data and specific 
parameterizations of these functions.

\begin{table}[h]
\caption{Comparison of the $|V_{ub}|$ extraction for $E_0=2.0\,\mathrm{GeV}$
for the two methods.
The first error reflects the uncertainty in the measurements 
of the $B\to X_u l\nu$ lepton spectrum, the second  error is due 
to the measurement of the $B\to X_S \gamma$ photon spectrum. 
For the shape-function based
analysis, the second error originates from the extraction of the shape-function 
parameters, in this case based on both the inclusive photon spectrum as well
as hadron-mass and lepton-energy moments from $B \to X_c l \nu$ decays. 
The third is the theoretical uncertainty. The fourth error for Method I
is due to the uncertainty of $|V_{ts}|$.
}
\begin{center}
\begin{tabular}{ll}
\hline\hline
Method  & $|V_{ub}| \cdot 10^3$ \\ \hline  
LLR     ~\cite{m3_llr1,m3_llr2} 
        & $4.28 \pm 0.29 \pm 0.29 \pm 0.26 \pm 0.28 $\\
LNP    ~\cite{lnp_2loop,lange}   
        & $4.40 \pm 0.30 \pm 0.41 \pm 0.23             $\\ \hline
SF-based analysis \cite{bbr_xulnu}  
        & $4.44 \pm 0.25 {\;}^{+0.42}_{-0.38} \pm 0.22$\\
\hline\hline
\end{tabular}
\end{center}
\label{t:res_20}
\end{table}

\begin{figure}[h]
\begin{center}
\includegraphics[height=5.8cm]{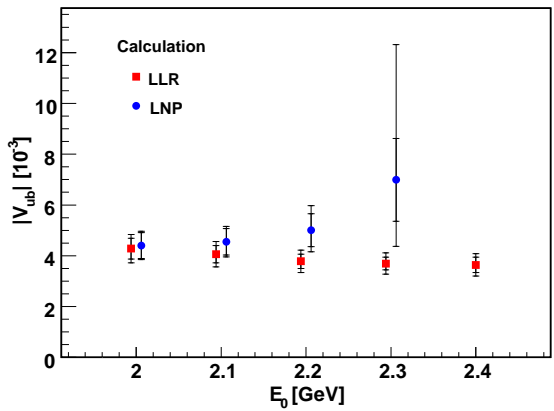}
\caption{
Comparison of $|V_{ub}|$ values extracted from the two different
calculations as a function of the lepton energy cut-off, $E_0$.
The errors bars represent the experimental and the total error.}
\label{f:compare}
\end{center}
\end{figure}

For comparison, the results for $E_0=2.0$~\gev for both methods are
presented in Table~\ref{t:res_20},
together with the \babar\ shape-function based measurement~\cite{bbr_xulnu}. 
Figure~\ref{f:compare} shows the dependence of the results on the lepton energy
cut-off, $E_0$.

The results agree well within their stated uncertainties.
Method's I results in values of $|V_{ub}|$ and errors that 
appear to be unaffected by the restriction of the 
phase space near the kinematic limit.
The theoretical errors 
are somewhat smaller than for Method II, and also smaller than 
for the previous extraction method, but there is an additional 
error due to the uncertainty of $|V_{ts}|$. 

The values of $|V_{ub}|$ extracted by using Method II for
the lepton energy cut-off at $E_0=2.0$\gev and $2.1\,\mathrm{GeV}$
agree well with those of Method I.   
The theoretical error contributions are estimated in detail. The smallest total
theoretical error is obtained for $E_0=2.0\,\mathrm{GeV}$;
it is about 5\%.  However, at larger $E_0$,
there are several increasing contributions to the theoretical 
uncertainties, in particular the first-order non-perturbative
hadronic power corrections that are not dealt with in Method I. 
This is why the error contribution from the weighted integral over
the photon spectrum increases very rapidly as the phase space
gets more restricted.

At present, for $E_0=2.0 $\gev, the experimental errors on both the
$B\to X_u \ell\nu$ branching fraction and the integral over $B\to X_s \gamma$
are about 12\%;  this means that there are opportunities to improve the
accuracy of $|V_{ub}|$ significantly by reducing the experimental errors
of these two spectra with more data available now and in the future. Also,
extending $E_0$ to 1.9~\gev or lower, would allow us to establish 
more clearly the stability of the results in this region. 

It would be of interest to perform a similar analysis for the hadron-mass
distribution in $B \to X_u \ell \nu$ decays, with increasingly tighter restrictions
on the mass of $X_u$, and assess whether this measurement is indeed more
robust than the measurement of the lepton spectrum near the kinematic endpoint.

\section{Acknowledgments}
                                                                                
This work has benefited from many interactions with our
theorist colleagues, Adam Leibovich and Ira Rothstein,
Matthias Neubert and Bjorn Lange. We would like
to thank Bjorn Lange for providing us with
the computer program for the numerical calculations for Method II.
We also are indebted to our \babar\ collaborators, in particular
members of the Analysis Working Group on Semileptonic Decays, above all
Jochen Dingfelder, Bob Kowalewski, Masahiro Morii, and Francesca di Lodovico.
This work has been supported by the U.S. Department of Energy and the Ministry of Science and Education
of Russian Federation. Two of us, V. G. and Yu. S., wish to thank SLAC for its support and kind hospitality.

\end{document}